\documentclass{article}

\usepackage{arxiv}

\usepackage[utf8]{inputenc} 
\usepackage[T1]{fontenc}    
\usepackage{hyperref}       
\usepackage{url}            
\usepackage{booktabs}       
\usepackage{amsfonts}       
\usepackage{nicefrac}       
\usepackage{microtype}      
\usepackage{lipsum}
\usepackage{graphicx}
\usepackage{amsmath}

\title{Cross-comparative analysis of evacuation behavior after earthquakes using mobile phone data}

\author{
  Takahiro Yabe \\
  Lyles School of Civil Engineering\\
  Purdue University\\
  West Lafayette, IN, USA \\
  \texttt{tyabe@purdue.edu} \\
   \And
  Yoshihide Sekimoto \\
  Institute of Industrial Science\\
  University of Tokyo\\
  Tokyo, Japan\\
   \AND
   Kota Tsubouchi \\
   Yahoo Japan Corporation \\
   Tokyo, Japan \\
   \And
   Satoshi Ikemoto \\
   Zenrin DataCom Corporation \\
   Tokyo, Japan \\
}

\begin{document}
\maketitle

\begin{abstract}
Despite the importance of predicting evacuation mobility dynamics after large scale disasters for effective first response and disaster relief, our general understanding of evacuation behavior remains limited because of the lack of empirical evidence on the evacuation movement of individuals across multiple disaster instances. 
Here we investigate the GPS trajectories of a total of more than 1 million anonymized mobile phone users whose positions are tracked for a period of 2 months before and after four of the major earthquakes that occurred in Japan. 
Through a cross comparative analysis between the four disaster instances, we find that in contrast with the assumed complexity of evacuation decision making mechanisms in crisis situations, the individuals’ evacuation probability is strongly dependent on the seismic intensity that they experience. 
In fact, we show that the evacuation probabilities in all earthquakes collapse into a similar pattern, with a critical threshold at around seismic intensity 5.5. 
This indicates that despite the diversity in the earthquakes profiles and urban characteristics, evacuation behavior is similarly dependent on seismic intensity. 
Moreover, we found that probability density functions of the distances that individuals evacuate are not dependent on seismic intensities that individuals experience.  
These insights from empirical analysis on evacuation from multiple earthquake instances using large scale mobility data contributes to a deeper understanding of how people react to earthquakes, and can potentially assist decision makers to simulate and predict the number of evacuees in urban areas with little computational time and cost, by using population density information and seismic intensity which can be observed instantaneously after the shock. 
\end{abstract}


\section*{Introduction}
Severe earthquakes such as the Kobe earthquake (1995), the Tohoku earthquake (2011) and more recently the Kumamoto earthquake (2016) led to mass evacuation activities owing to considerable damage to buildings and urban infrastructure (\cite{yun2015evacuation,heath1995kobe,mimura2011damage}). 
Human mobility prediction in disaster scenarios is crucial for various recovery efforts including planning the locations and capacities of evacuation shelters and the allocation of various emergency supplies. 
Many of the conventional methods use urban infrastructure failure data to predict the number of evacuees in upcoming disasters. 
Tokyo Metropolitan Government uses a model composed of variables such as building collapse rate, lifeline damage rate, and inundation area rate to estimate the number of evacuees (\cite{tokyometro}). 
However, it is difficult to utilize this model shortly after an earthquake because typically, several days are required for government organizations to inspect and gather information about the status of lifelines and infrastructure.
In fact, after the Kumamoto earthquake (2016, magnitude 7.3), authorities failed to obtain an accurate estimate of the number of evacuees owing to difficulties in information gathering activities. 
This caused delay in rescue and inefficient distribution of emergency supplies (\cite{Kumamoto}). 

Traditionally, transportation surveys were used to understand the city-scale human mobility (\cite{sekimoto2011pflow}).
In recent years, large scale datasets collected from mobile phones and smartphones are beginning to be utilized to understand the behavior of individuals at a low cost (\cite{phithakkitnukoon2012socio,jiang2016timegeo,wardrop2018spatially,gonzalez2008understanding,calabrese2011estimating,santi}). 
These data are utilized in applications in several fields such as traffic management (\cite{iqbal2014development,wang2012understanding,demissie2013intelligent}), monitoring pandemic spreading  (\cite{bengtsson2011improved}), tourist mobility analysis (\cite{phithakkitnukoon2015understanding}) and the prediction of population distributions and dynamics at various scales (\cite{deville2014dynamic,shimosaka2015forecasting,nishi2014hourly,noulas2012tale}). 
Many works have applied this new data source for applications in disaster management (\cite{gething2011can}).
Studies have shown the effect of weather patterns on human mobility (\cite{horanont2013weather}), and its predictability using socio-economic factors (\cite{yabe2016predicting}).
In terms of understanding human mobility during larger scale disasters, Lu \textit{et al.} analyzed call detail records to investigate the predictability of evacuation destinations of individuals after the Haiti earthquake and showed that most evacuation destinations were places that individuals has visited earlier (e.g., the home of a relative or a friend) (\cite{lu2012predictability}).
Using Twitter Geo-tag data, Wang \textit{et al.} showed that the radius of gyration of the victims of Hurricane Sandy changed significantly from their typical behavior (\cite{wang2014quantifying}). 
Song \textit{et al.} analyzed the long-term evacuation behavior of the victims of the Tohoku earthquake and clarified the population decline in various affected areas (\cite{song2013intelligent}). 
Other works have used Twitter data to understand the tweeting behaviors of affected individuals and the transition of sentiments after large scale disasters (\cite{kryvasheyeu2015performance,kryvasheyeu2016rapid}).
More recent studies conducted after the Nepal earthquake and Kumamoto earthquake showed that evacuation behavior could be monitored after an earthquake by using the mobile phone location data (e.g. GPS, call detail records) obtained from the evacuees (\cite{horanont2013large,yabe2016estimating,wilson2016rapid}). 
Although these works provide valuable insights into the human mobility patterns during a individual disaster case studies, they fail to provide general insights that could be applicable across different disasters. 
Moreover, results obtained from mobile phone data are usually delivered to decision makers several days or even weeks after the initial shocks because the use of real-time mobile phone location data is highly sensitive against privacy issues (\cite{de2013unique}). 
This motivates us to perform a cross-comparative analysis of human evacuation behavior after various disaster cases, to obtain a general understanding of evacuation behavior after earthquakes which can be used in planning evacuation strategies for future disasters.

Several works have performed a cross-comparative analysis across different disasters.
However, to the best of our knowledge, none of the works provide analyses of detailed evacuation behavior which can be utilized by emergency management practitioners (\cite{bagrow2011collective,wang2016patterns}). 
Bagrow \textit{et al.} studied the calling behavior after various types of disasters and found that communication spikes after emergencies are both spatially and temporally localized, but information about emergencies spreads globally (\cite{bagrow2011collective}). 
Here in this study, we focus more on the physical movements of individuals rather than communication patterns. 
Wang et al. study the radius of gyration of individuals across different disaster cases using Twitter Geo-tag data (\cite{wang2016patterns}). 
Although using the radius of gyration measures the perturbation of human mobility due to disasters, it does not provide direct measures of evacuation movements. 
In this study, we provide detailed analysis of evacuation rates and evacuation distances of individuals after multiple earthquakes using large scale mobile phone location datasets from Japan. 
More specifically, we analyze the mobility of individuals who were affected by the Tohoku earthquake (2011 March), Kumamoto earthquake (2016 April), Nagano earthquake (2014 November), and Tottori earthquake (2016 October).


\section*{Data}
\subsection*{Mobile phone location data}
Yahoo Japan Corporation\footnote{https://about.yahoo.co.jp/info/en/} provides a variety of disaster notifications to users through its disaster prevention application. 
The disaster prevention application continues to acquire real-time location information of individuals in order to transmit only relevant disaster notifications for individuals. 
The users have accepted to provide their location data when installing the application. 
The data is anonymized so that individuals cannot be specified, and personal information such as gender, age and occupation are unknown. Each location data is stored as a GPS record in Yahoo! Japan's internal server. 
Each record consists of a user's unique ID (random character string), latitude, longitude, date and time. 
The acquisition frequency of the GPS data changes according to the movement speed of the user. 
If it is determined that the user is staying in a certain place for a long time, data is acquired at a relatively low frequency, and if it is determined that the user is moving, the data is acquired more frequently. 
By reducing the number of times data is acquired using this algorithm, it is possible to reduce the burden on the user's smartphone. 
On average, about 40 points are observed per day per user, therefore we can observe the main staying places of each individual. 
Currently the number of users who have installed the disaster alert app in Japan nationwide are about 1 million people (Kumamoto, at 2016 April), and it is currently increasing due to the rise in disaster risk awareness. 
Approximately, this is 1\% sample rate of the whole population which is equivalent to the national census, which is performed once in 10 years for the purpose of understanding traffic behavior. 
\nameref{S1_Kumamotoplot} shows the actual GPS information for one day plotted on a white map around Kumamoto City. Each fine red dot corresponds to one GPS data. 
We can observe that the center of the city and even the roads that connect the satellite cities can be seen from the data, showing its richness. 
Table \ref{stats} shows the statistics of the disaster and also the data that were used for the experiments. 
We used 712,904 users' data for which an average of 19 points were observed per day for the Kumamoto earthquake. 
For Nagano and Tottori earthquakes, the number of users were 10244 and 24103, respectively with similar observation frequencies. 
We also use the GPS data set (called Konzatsu-Tokei (R) data), provided by Zenrin Data Com\footnote{https://www.zenrin-datacom.net/en/index.html} (ZDC) to anlyze the evacuation mobility during the Tohoku earthquake.
“Konzatsu-Tokei (R)" Data refers to people flows data collected by individual location data sent from mobile phone with enabled AUTO-GPS function under users' consent, through the “docomo map navi" service provided by NTT DOCOMO, INC. 
Those data is processed collectively and statistically in order to conceal the private information. 
Original location data is GPS data (latitude, longitude) sent in about every a minimum period of 5 minutes and does not include the information to specify individual such as gender or age. 
To analyze the evacuation behaviors of people at the time of the Tohoku earthquake, we used the data from February 1, 2011 to March 31, 2011, and the total number of IDs in the Tohoku region (Aomori, Iwate, Akita, Miyagi, and Yamagata prefectures) was 157,225. 
To analyze the evacuation behavior due to the effects of earthquakes and to separate the effects of other disaster types, we excluded areas that were hit by the tsunami (coastal areas) and cities in Fukushima, which was heavily affected by the nuclear power plant accident, as shown in Fig \ref{fig1}.  

\begin{table}
\centering
\caption{
{\bf Statistics of the four earthquakes and mobile phone data used for analysis}}
\begin{tabular}{l|rcc|cc}
\hline
 & \multicolumn{3}{c|}{Disaster Statistics} & \multicolumn{2}{c}{Data Statistics} \\
Disaster & \multicolumn{1}{c}{Date} & Max. SI & No. LGUs (SI$\geq$4.0) & Obs. Period & No. Users \\
\hline
Tohoku Earthquake   & 2011/3/11 & 6.3 & 140 & 2011/02/01 $\sim$ 03/31 & 157,225  \\
Kumamoto Earthquake & 2016/4/16 & 6.7 & 153 & 2016/03/15 $\sim$ 05/15 & 712,901  \\
Tottori Earthquake  & 2016/10/21 & 5.7 & 19 & 2016/10/01 $\sim$ 11/30  & 24,103  \\
Nagano  Earthquake  & 2014/11/22 & 5.7 & 10 & 2014/11/01 $\sim$ 12/30 & 10,244  \\ \hline
\end{tabular}
\label{stats}
\end{table}

\subsection*{Seismic intensity data}
\begin{figure}
\includegraphics[width=\textwidth]{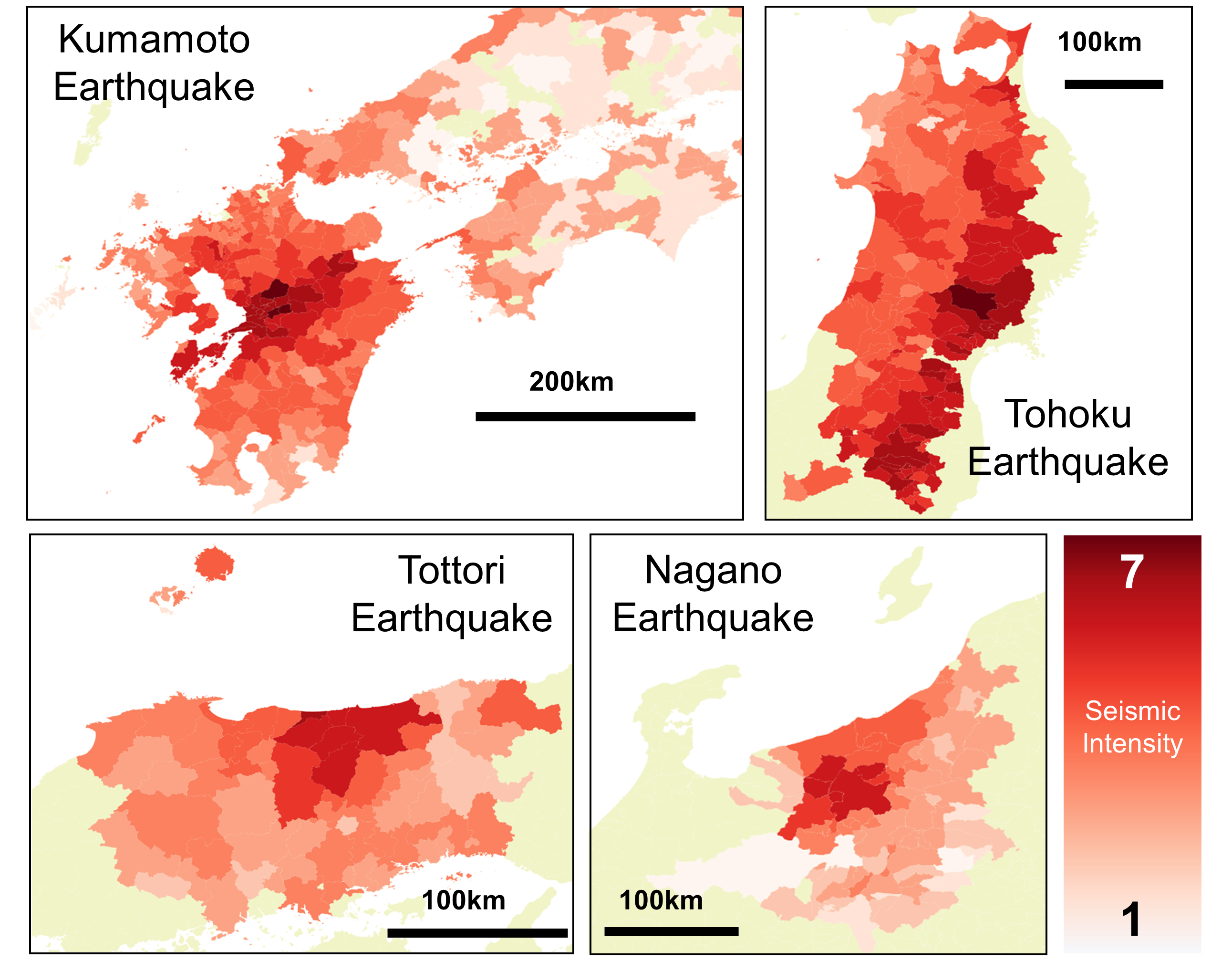}
\caption{{\bf Areas that were analyzed in the four earthquakes.}
Seismic intensity values observed in each municipality is shown by the darkness of red colors. Note that for the Tohoku Earthquake, the coastal areas in Iwate and Miyagi prefectures and areas in Fukushima prefecture that were affected by the nuclear power plant accident was removed from the analysis.}
\label{fig1}
\end{figure}

"Seismic intensity" is used as an index to measure the strength of the external force of earthquakes, varying from 1 (very small) to 7 (extreme shock), with 0.1 increments.
Machiki City experienced the largest seismic intensity of 6.7 during the Kumamoto earthquake.
Seismic intensity is obtained from the acceleration caused by the earthquake, and each local government unit is given a seismic intensity value.  
The seismic intensity data is published in the “Jishin/Kazan Geppou (Monthly magazine of earthquakes and volcanoes)” published by the Japan Meteorological Agency. 
Note that seismic intensity is different from ``magnitude'', which is only observed at the epicenter. 
We used the data from 2016 April and October, 2014 November, and 2011 March. Fig \ref{fig1} shows the seismic intensity data for each local government unit (LGU). 

\section*{Data Analysis and Results}

\subsection*{Home location estimation and evacuation}
It is well known that human trajectories show a high degree of temporal and spatial regularity, each individual having a significant probability to return to a few highly frequented locations, including his/her home location (\cite{gonzalez2008understanding}).
Due to this characteristic, it has been shown that home locations of individuals can be detected with high accuracy by clustering the individual's stay point locations over night (\cite{calabrese2011estimating}).
Home locations of each individual was detected by applying mean-shift clustering to the nighttime staypoints (observed between 8PM and 6AM), weighted by the duration of stays in each location (\cite{ashbrook2003using,kanasugi2013spatiotemporal}).
Mean shift clustering was implemented using the scikit-learn package on Python\footnote{http://scikit-learn.org/stable/modules/generated/sklearn.cluster.MeanShift.html}.
An individual was detected to be evacuated if the individual is estimated to be staying more than $r$ (meters) away from his or her estimated home location. 
Evacuation rate on day for a given city is calculated by dividing the number of evacuated individuals by the total number of users that were estimated to be living in that city.
For each disaster $d$, we calculate the evacuation rate $p_d(z)$ at a given seismic intensity $z$ by the following equation. 
\begin{equation}
    p_d(z) = \frac{\sum_{i\in S_d(z)} M^*_i}{\sum_{i\in S_d(z)} M_i}
\end{equation}
where $S_d(z)$ is the set of LGUs that experienced a seismic intensity of $z$ in disaster $d$, $M_i$ is the total number of users living in LGU $i$, and $M_i^*$ is the number of evacuated people from LGU $i$, which we observe from trajectories of mobile phone location data.

\subsection*{Accuracy of mobile phone location data}
The GPS data provided by Yahoo! Japan has a sample rate of about 1\%, which is about the same as the sample rate of the personal trip survey conducted by the Ministry of Land, Infrastructure and Transport, so it is suggested that this sample rate is enough for grasping the entire urban flow. 
Moreover, Deville \textit{et al.} shows that the distribution of the entire population can be accurately reproduced from the call detail record data obtained from mobile phones (\cite{deville2014dynamic}). 
To show that our data is also sufficient to represent the whole population, we examine how accurate the actual population distribution can be estimated from the nighttime GPS data. As verification data, we use the national census data, which contains the residential population data for every 1,000 meter grid.
We separate Kumamoto Prefecture into 1,000 meter grid, and we estimate the population in each grid mesh by multiplying the number of IDs in the grid mesh from the GPS dataset by the inverse of the sample rate. 
The correlation coefficient between the estimated population and the actual population is 0.853, showing a high accuracy of population estimation (\nameref{correlation}).

\subsection*{Spatial analysis of seismic intensity and evacuation}
\begin{figure}
\includegraphics[width=\textwidth]{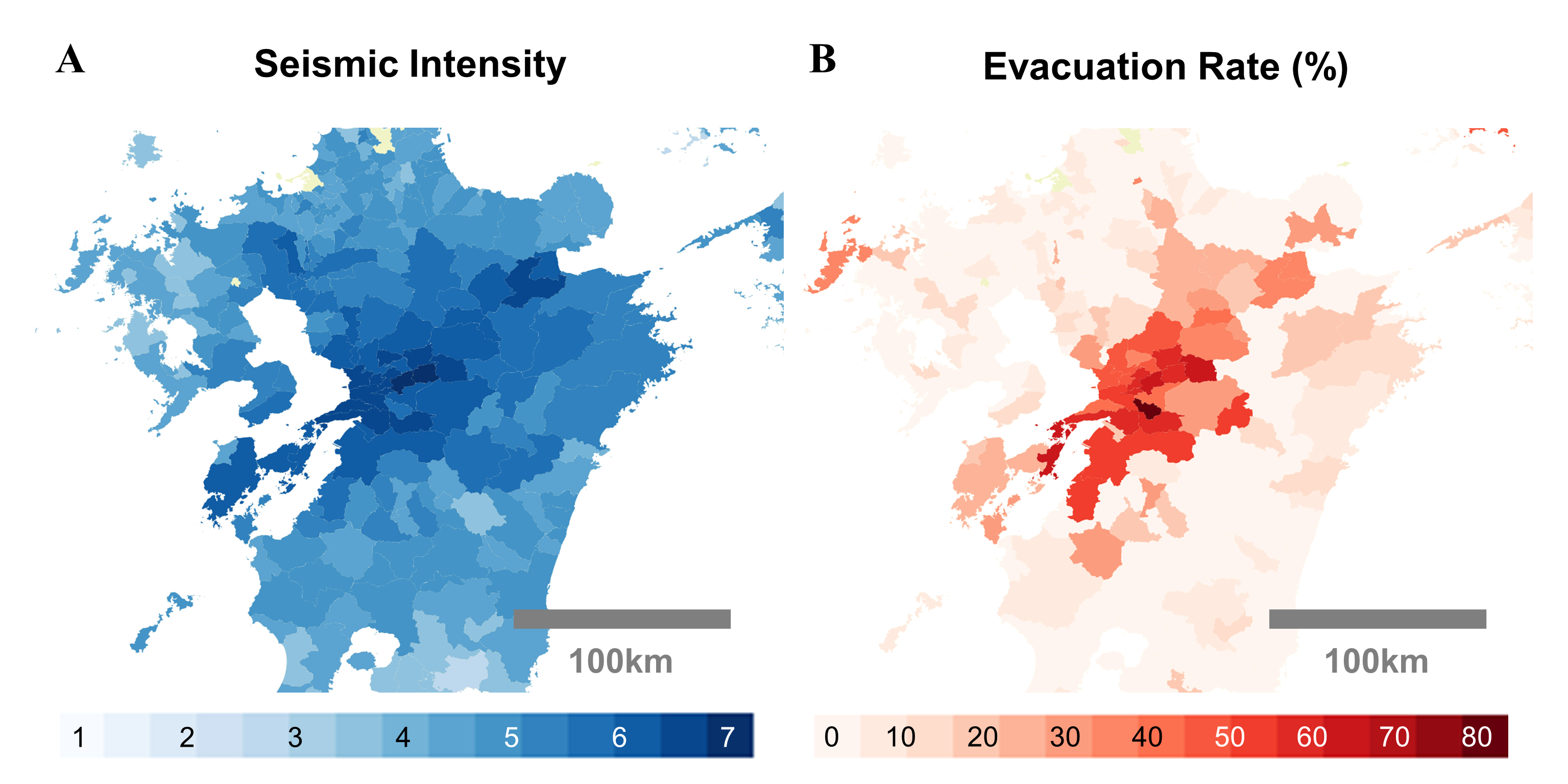}
\caption{{\bf Spatial distribution of seismic intensity and evacuation rates}
A: Seismic intensity data shown for each local government unit (LGU). B: Evacuation rate for each LGU. Seismic intensity and evacuation rates are positively correlated, however evacuation rates suddenly increase at around 5.}
\label{fig2}
\end{figure}

Fig \ref{fig2} shows the spatial distribution of seismic intensity (A) and evacuation rate (B) after the Kumamoto earthquake. 
Both data are aggregated into local government unit (LGU) areas. 
The evacuation rate in a given LGU is calculated by dividing the estimated number of individuals who are staying more than 200 meters away from their estimated home location after the earthquake by the total number of individuals who are estimated to be living in the given LGU. 
We can observe that even though seismic intensity gradually attenuates as the distance from the epicenter increases, evacuation rate abruptly decreases from approximately 40\% to less than 10\% around areas with a seismic intensity of 5.0. 
We can infer that human evacuation activities have a critical ‘tipping point’ of seismic intensity. 
The underlying reasons for this behavior are not trivial.
However, it can be assumed that evacuation activities increase sharply owing to the collapse of buildings and critical social infrastructure such as electricity and gas lines, which occur at similar seismic intensities. 
We model this nonlinear characteristic of evacuation activities by further observing the evacuation activities for four large earthquakes in Japan. 

\subsection*{Fragility curve for evacuation behavior}
\begin{figure}
\includegraphics[width=\textwidth]{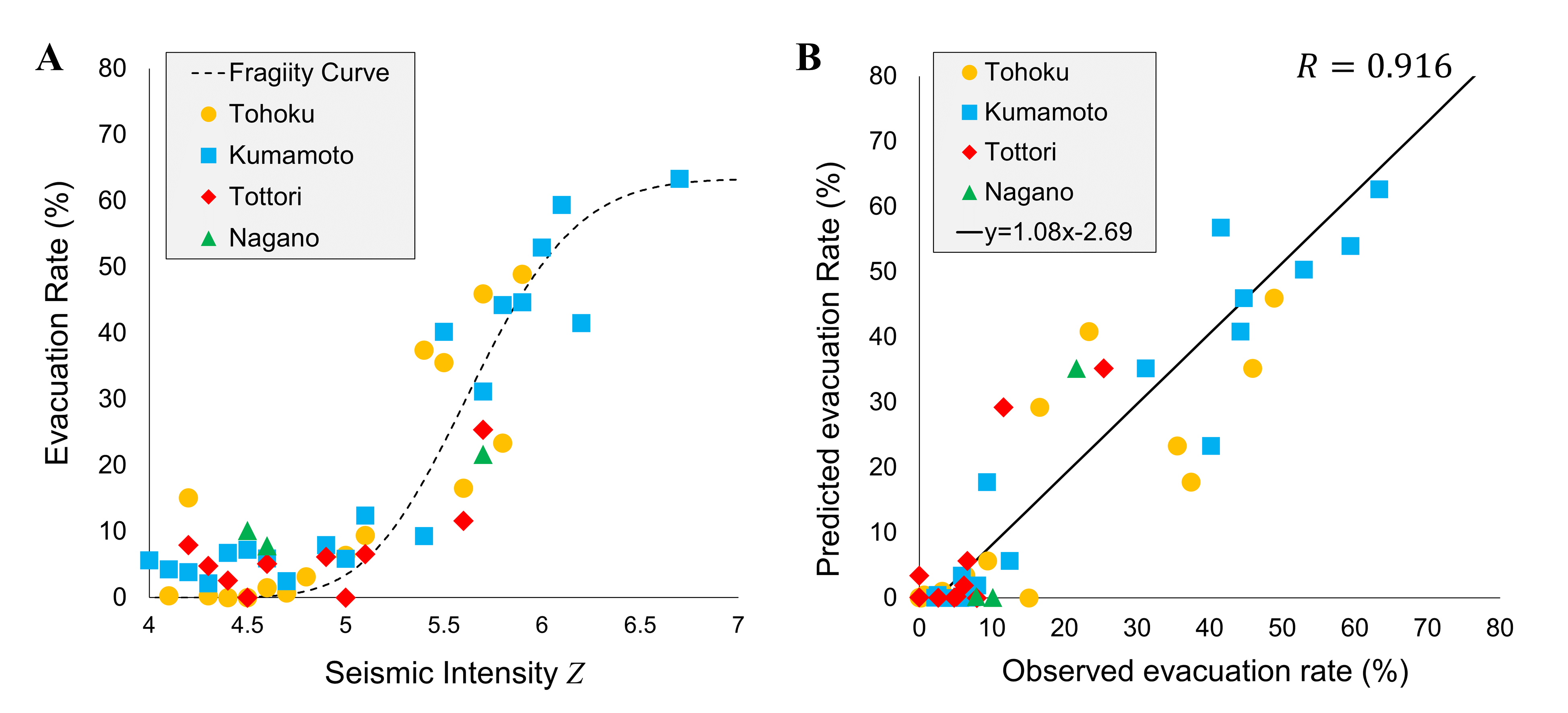}
\caption{{\bf Fragility curves for evacuation rates.}
A: Evacuation rates for each of the four earthquakes plotted against seismic intensities for all affected LGUs, along with the estimated fragility curve. Colors indicate earthquake incident. B: Scatter plot of estimated and true evacuation rates.}
\label{fig3}
\end{figure}

To explore the general properties of the evacuation activities after large earthquakes, we analyzed each individual’s movement after the initial shocks of the four earthquakes. 
Out of these earthquakes, the Tohoku earthquake caused multiple types of hazards including a tsunami and a nuclear power plant accident. 
In this study, we focus on areas that were affected only by the earthquake to understand the evacuation behavior caused by large earthquakes. 
Therefore, the LGUs in the Fukushima prefecture and the LGUs along the coastal line in the Tohoku region are out of the scope of this study ({tsubokura2013limited}. 
For other earthquakes, we calculated the evacuation rate of all LGUs that experienced a seismic intensity of larger than 4.0.
Fig \ref{fig3}A shows the evacuation rates during the four earthquakes according to their seismic intensities. The four different colors and plots correspond to the different earthquakes. 
To model the sudden increase of evacuation rates with respect to seismic intensities, we use the fragility curve model.

The fragility curve is a model used in the field of structural mechanics, mainly for modelling the collapse rate of a building with the seismic intensities of earthquakes and the inundation height of tsunami as variables. 
As shown in Koshimura \textit{et al.}, the external force of the disaster is taken on the horizontal axis and the vertical axis is the collapse rate. 
Koshimura \textit{et al.} discovered that the inundation depth of the tsunami and the collapse rate of the building follow the cumulative log normal distribution function (\cite{koshimura2009developing}). 
Similarly, Baker \textit{et al.} found that the seismic intensity of the earthquake and the collapse rate of the building can also be approximated with the cumulative log normal distribution function (\cite{baker2015efficient}). 
The common feature of both discoveries is that the collapse rate of buildings is close to 0 up to a certain external force but rapidly increases from a certain inflection point. 
The cumulative log normal distribution is a function that can model this sudden increase. 
The cumulative log normal distribution function is a function shown in equation, and there are three parameters ($\mu$, $\sigma$, $a$). 
$\mu$ and $\sigma$ dictate the slope of the function and the position of the inflection point, while $a$ determines the maximum value of evacuation rates. 
We model the evacuation rate of humans $p(z)$ given seismic intensity $z$ as the following fragility function, and estimate parameters $\hat{\mu},\hat{\sigma},\hat{a}$ using the maximum likelihood estimation approach (\cite{baker2015efficient}) given the observations from mobile phone data. 
\begin{equation}
    p(z) = a \Phi(\frac{\ln{z}-\mu}{\sigma}) = a \int_0^z \frac{1}{\sqrt{2 \pi} \sigma} \exp{\big( \frac{- (\ln{t} -\mu)^2}{2 \sigma^2} \big)} dt
    \label{pz}
\end{equation}

The estimated parameters of the fragility curve using the four disaster cases are $\mu=1.73$, $\sigma=0.075$, $a=0.63$, and the correlation coefficient between the model and the data was $R=0.916$, as shown in Fig \ref{fig3}B.
From these estimations, we can infer that the seismic intensity at which individuals start to evacuate is approximately 5.2 and evacuation rate increases sharply between intensities of 5.5 and 6.0. 
At a seismic intensity of approximately 6.5, evacuation rate plateaus at 63\%.
This plateau explains the strength of the infrastructure in Japan because it implies that regardless of seismic intensity, a relatively large fraction (approximately 34\%) of the individuals do not have to evacuate from their homes.
We test the robustness of our results by performing a leave-one-out test similar to cross validation but with stricter methods. 
More specifically, we estimate the parameters of the fragility curve using data obtained from three disasters, and then test the fit for the left out disaster by measuring the correlation coefficient $R$ and mean average percentage error (MAPE) of evacuation rates.
Table \ref{robust} shows the robustness testing results. 
The first row means that the parameters of the fragility curve were estimated using data from disasters except Tohoku earthquake (i.e. Kumamoto, Tottori, Nagano earthquakes) as $\hat{\mu}=1.79$ and $\hat{\sigma}=0.122$, and were used to predict the evacuation rates after the Tohoku earthquake, which showed high accuracy ($R=0.879$, MAPE=$5.35\%$). 
For all disasters, prediction showed high correlation and small MAPE, all below 10\%. 
Moreover, the estimated parameters were similar for all testing cases, showing the high stability and robustness of the estimated fragility curve.
This test shows that by using the fragility curve fitted by data from past disasters, we are able to predict the evacuation rates in future disasters with high precision.


\begin{table}
\centering
\caption{
{\bf Robustness test of fragility curve fitting results. The evacuation rates for the ``left out disaster'' was predicted by the model fitted by the other three disasters.}}
\begin{tabular}{l|cc|cc}
\hline
\textbf{Disaster left out} & \multicolumn{2}{c|}{\bf Prediction Performance} & \multicolumn{2}{c}{\bf Estimated parameters} \\
\textbf{for prediction} & R     & MAPE (\%)  & $\Hat{\mu}$   & $\Hat{\sigma}$ \\
\hline
Tohoku            & 0.879 & 5.35 & 1.79 & 0.122 \\
Kumamoto          & 0.944 & 7.04 & 1.86 & 0.185 \\
Tottori           & 0.822 & 4.92 & 1.81 & 0.144 \\
Nagano            & 0.984 & 7.91 & 1.80 & 0.137 \\ \hline
\end{tabular}
\label{robust}
\end{table}

\begin{figure}
\includegraphics[width=\columnwidth]{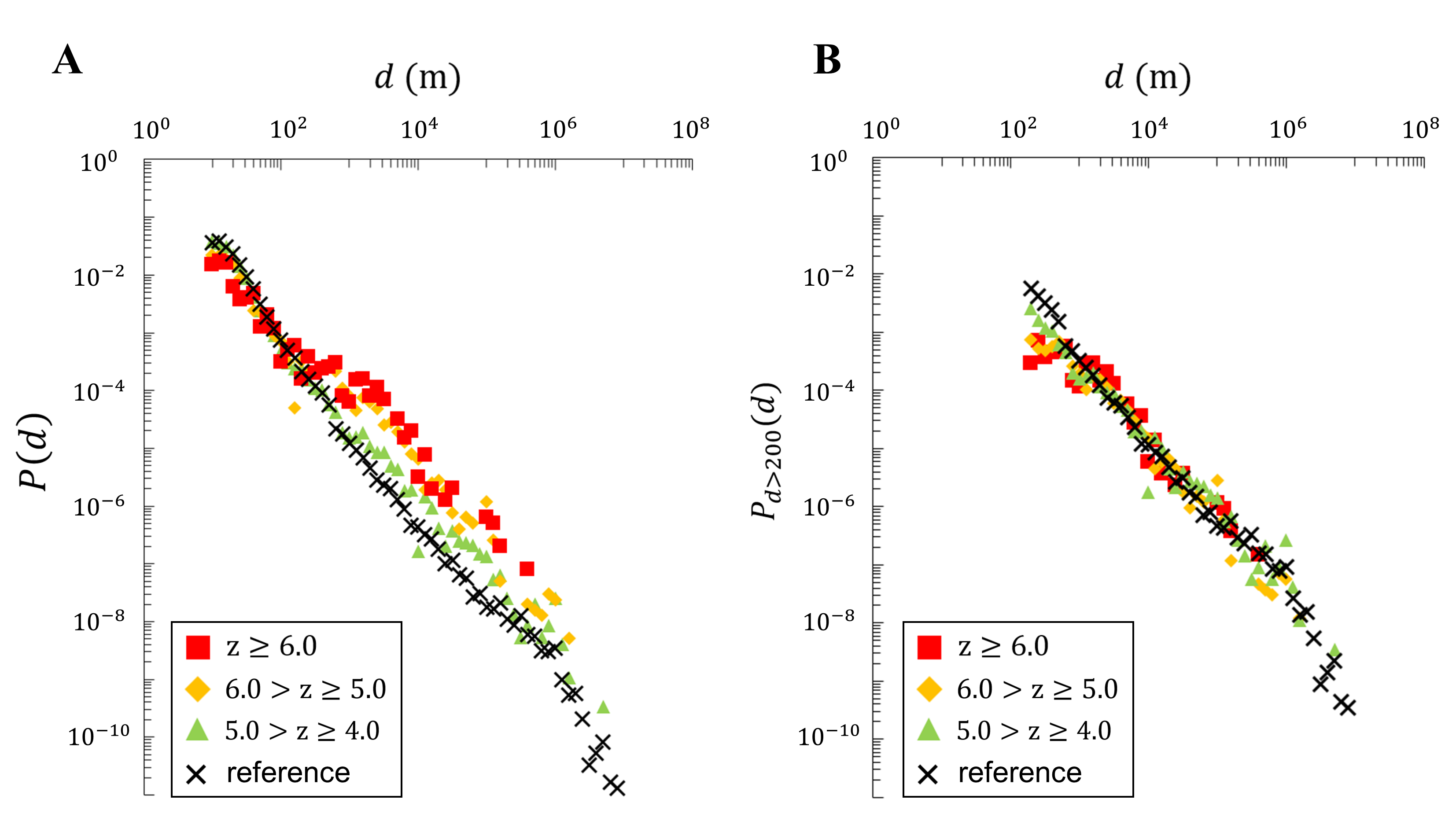}
\caption{{\bf Distribution of evacuation distance.}
A: Distribution of evacuation distances including all observed users. B: Distribution of evacuation distances for only those who evacuated. We can observe that the probability densities for different seismic intensities collapse into one distribution, implying that evacuation distances are not dependent on seismic intensity.}
\label{fig4}
\end{figure}

Understanding the spatial and temporal characteristics of evacuation behavior is important for creating effective evacuation plans. 
We investigated the evacuation distance, which is shown in Fig \ref{fig4}.
The long tail distribution of the evacuation distance follows a power law $P(d)=\alpha d^{-\gamma}$, similar to cases in non-disaster cases (\cite{gonzalez2008understanding}). 
We can observe the evacuation distance to be between 1 kilometer to 1000 kilometers for all levels of seismic intensities. 
Moreover, when we cutoff the individuals who did not evacuate and plot the evacuation distance distributions for only the evacuated individuals (i.e. $d>200$), we can observe that the distance distribution collapse into one distribution despite different seismic intensities with $1.13 \leq \gamma \leq 1.38$. 
This implies that if individuals decide to evacuate from their homes, their destination is not affected by the intensity of the earthquake. 
Moreover, this shows that individuals have limited options for destinations, such as evacuation shelters and the houses of their relatives, which reinforces the findings from Lu \textit{et al.} (\cite{lu2012predictability}).
Through our cross-comparative analysis, we were able to expand their findings to a general setting.  

\section*{Conclusion}
Using large scale mobility data collected from over 1 million mobile phones of users affected by earthquakes, we carried out a cross-comparative analysis on the evacuation mobility patterns of individuals. 
Through a cross comparison across four large scale earthquakes in Japan, we found that although city characteristics vary, the evacuation rates can be approximated well with a single fragility curve with respect to seismic intensities, with a high correlation of $R=0.916$. 
The robustness of the fragility curve was checked by performing a cross-validation test on the four disasters. 
Moreover, it was found that the distribution of evacuation distances did not depend on the seismic intensity that the individual experienced, which extends the past findings to a general setting. 
Our data-driven analysis of evacuation behavior after earthquakes based on seismic intensity could improve the manner in which disaster managers prepare and respond to disasters. 
Since seismic intensity data can be obtained instantly after the shock, practitioners can use that information to predict the approximate number of evacuees instantaneously after the earthquake to develop evacuation shelter location plans and significantly improve the quality of disaster response. 
Our future works include analyzing the fragility curves of evacuation rates in different areas of the world apart from Japan to investigate the factors that characterize robustness of cities against earthquakes. 

\section*{Supporting information}

\paragraph*{S1 Fig.}
\label{S1_Kumamotoplot}
{\bf Mobile phone location data} GPS data obtained during 1 day plotted onto a white map in Kumamoto area. The mobile phone data is dense in both spatial and temporal aspects to analyze the detailed mobility of individuals.
\begin{figure}
\includegraphics[width=0.6\textwidth]{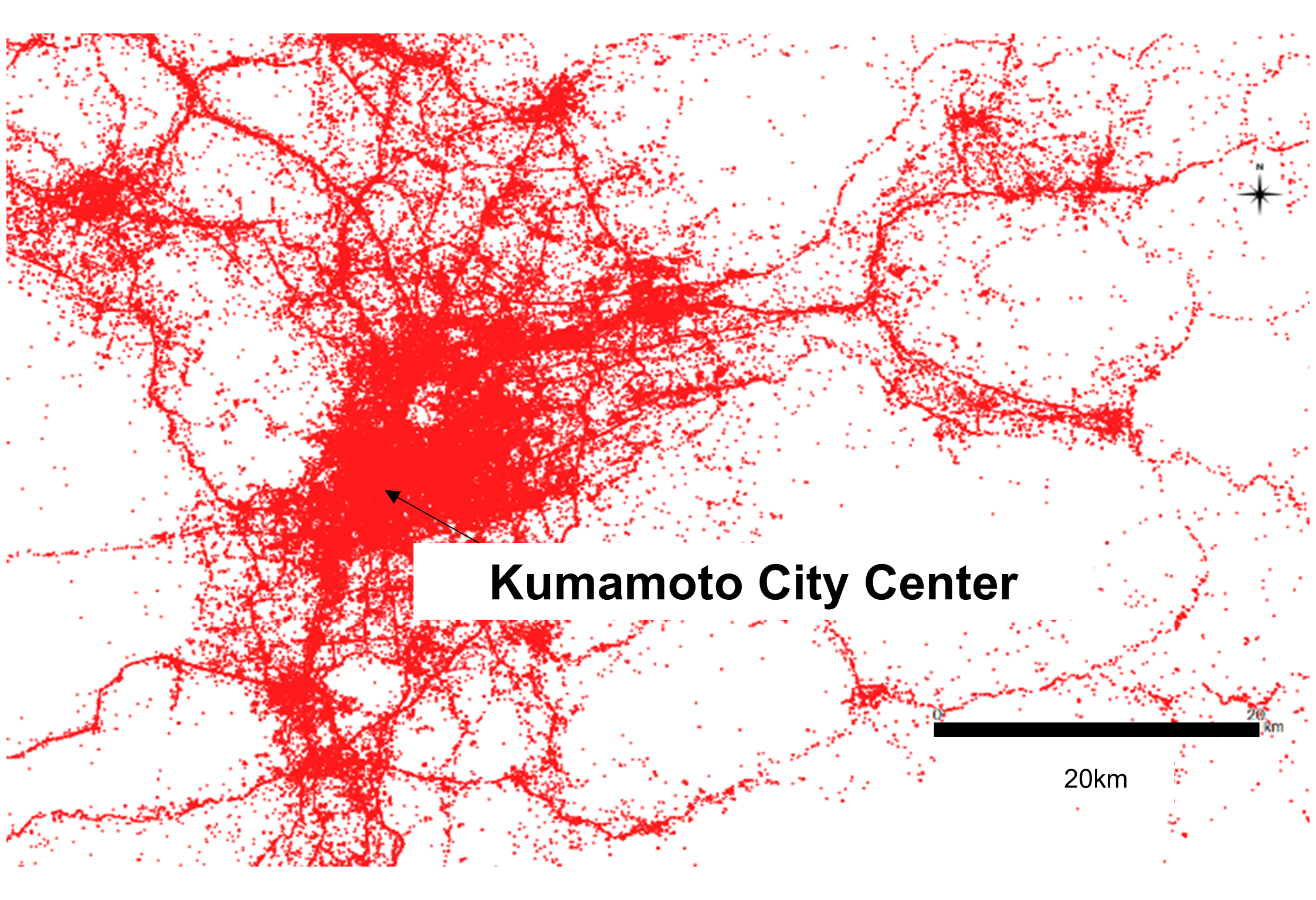}
\label{figS1}
\end{figure}

\paragraph*{S2 Fig.}
\label{correlation}
{\bf Comparison between national census data and population density estimated from GPS data.}
Estimated population and the population obtained from the census. 
The blue dots correspond to one grid mesh (1000m size) respectively.
\begin{figure}
\includegraphics[width=0.7\textwidth]{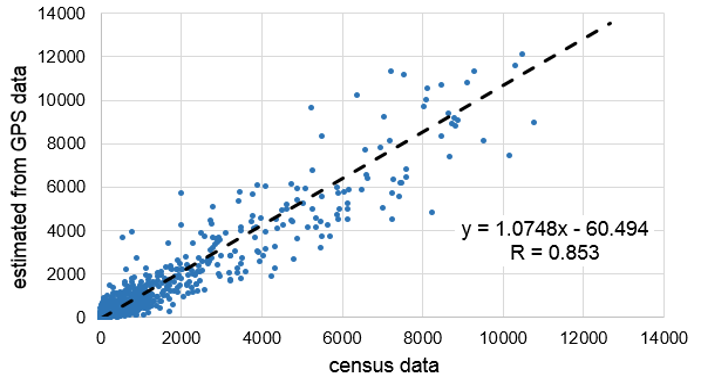}
\label{figS2}
\end{figure}

\paragraph*{S3 Fig.}
\label{sensi}
{\bf Sensitivity analysis on $r$.}
Fragility curves with different $r$ parameter values. Although the estimated parameters vary under different parameters, the general findings are not affected, where the fragility curves fit the result well. 
\begin{figure}
\includegraphics[width=0.6\textwidth]{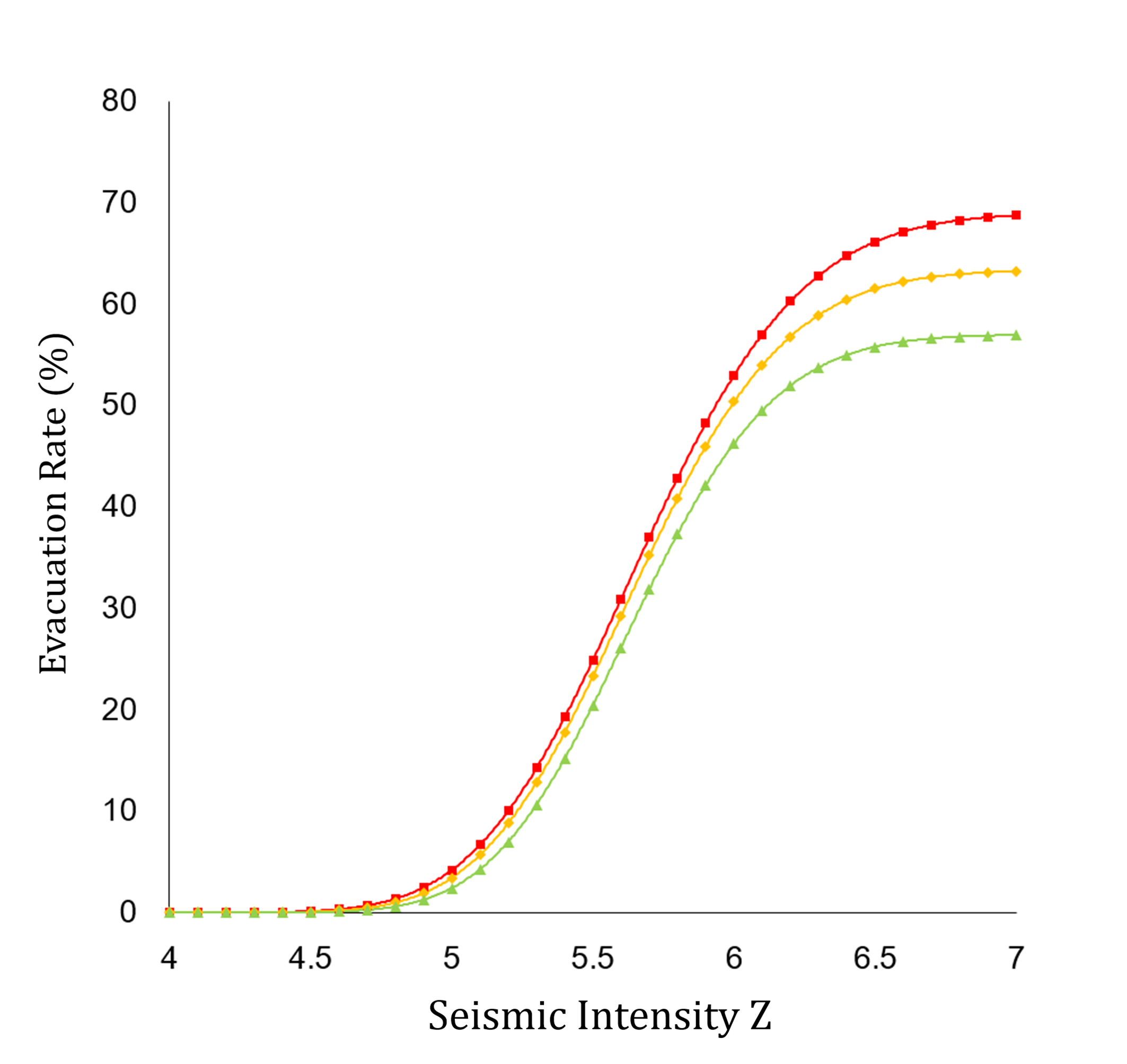}
\label{figS3}
\end{figure}

\paragraph*{S4 Fig.}
\label{timing}
{\bf Evacuation timing after earthquake.}
Evacuation timing of individuals after the Kumamoto earthquake. Results show that the higher the seismic intensity, individuals evacuate more quickly.
\begin{figure}
\includegraphics[width=\textwidth]{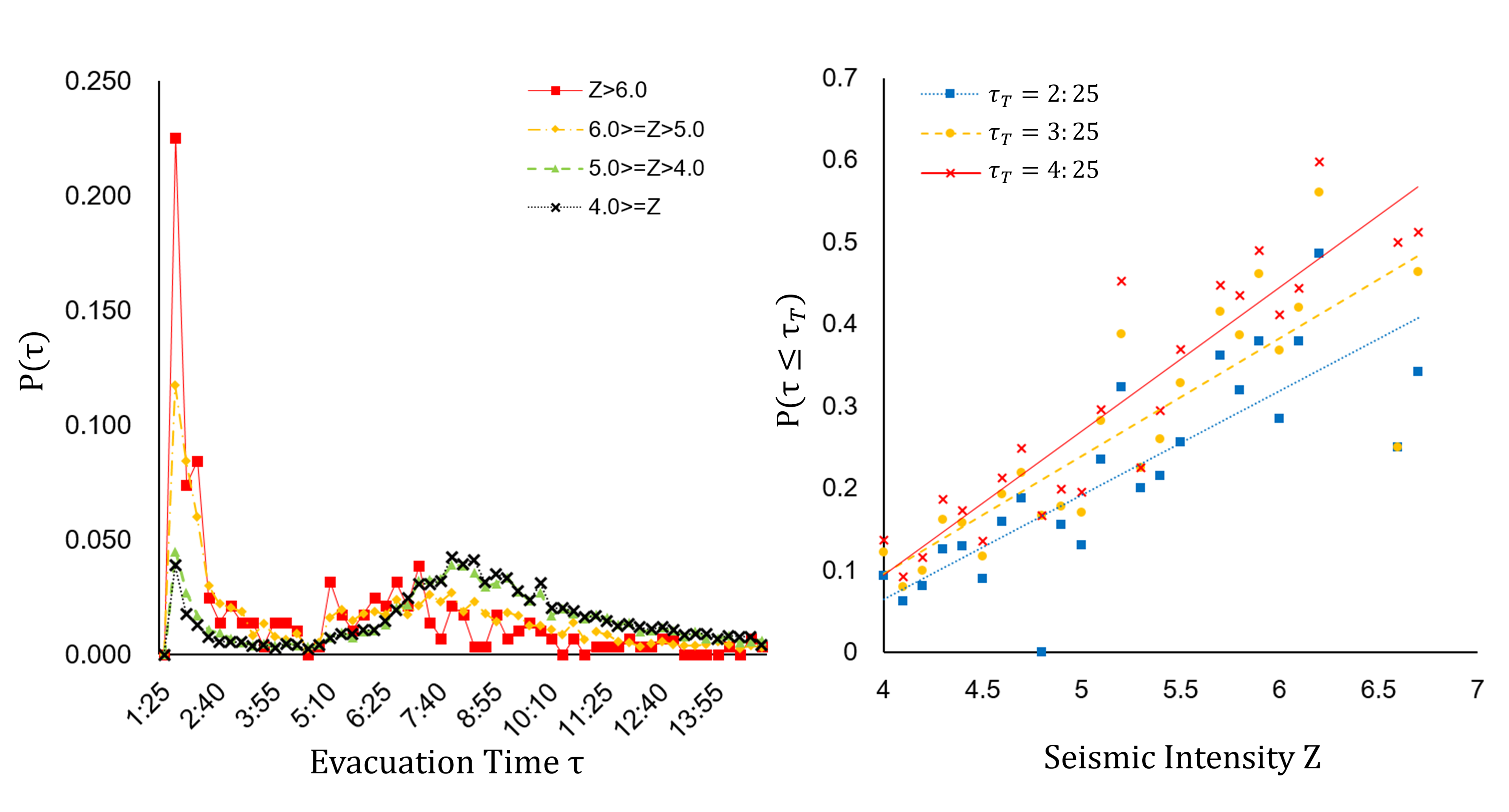}
\label{figS4}
\end{figure}

\section*{Acknowledgments}
We thank Dr. Naoya Fujiwara of Tohoku University and Dr. Akihito Sudo of Shizuoka University for many fruitful discussions.
Takahiro Yabe was supported by Grant-in-Aid for Japan Society for the Promotion of Science (JSPS) Research Fellow 17J09460.

\section*{Author Contributions}
T.Y., Y.S., and K.T. designed research; T.Y. and K.T. performed research; T.Y. and K.T. analyzed data; T.Y., Y.S., K.T., and S.I. wrote the paper.

\bibliographystyle{plainnat}
\bibliography{references}

\end{document}